# Life Before Earth

Alexei A. Sharov, Ph.D.
Staff Scientist, Laboratory of Genetics
National Institute on Aging (NIA/NIH)
333 Cassell Drive Baltimore, MD 21224 USA
sharoval@mail.nih.gov

Richard Gordon, Ph.D.
Theoretical Biologist, Embryogenesis Center
Gulf Specimen Marine Laboratory, P.O. Box 237, 222 Clark Drive
Panacea FL 32346 USA
DickGordonCan@gmail.com

**Abstract**

An extrapolation of the genetic complexity of organisms to earlier times suggests that life began before the Earth was formed. Life may have started from systems with single heritable elements that are functionally equivalent to a nucleotide. The genetic complexity, roughly measured by the number of non-redundant functional nucleotides, is expected to have grown exponentially due to several positive feedback factors: (1) gene cooperation, (2) duplication of genes with their subsequent specialization (e.g., via expanding differentiation trees in multicellular organisms), and (3) emergence of novel functional niches associated with existing genes. Linear regression of genetic complexity (on a log scale) extrapolated back to just one base pair suggests the time of the origin of life = $9.7 \pm 2.5$ billion years ago. Adjustments for potential hyperexponential effects would push the projected origin of life even further back in time, close to the origin of our galaxy and the universe itself, 13.75 billion years ago. This cosmic time scale for the evolution of life has important consequences: (1) life took a long time (ca. 5 billion years) to reach the complexity of bacteria; (2) the environments in which life originated and evolved to the prokaryote stage may have been quite different from those envisaged on Earth; (3) there was no intelligent life in our universe prior to the origin of Earth, thus Earth could not have been deliberately seeded with life by intelligent aliens; (4) Earth was seeded by panspermia; (5) experimental replication of the origin of life from scratch may have to emulate many cumulative rare events; and (6) the Drake equation for guesstimating the number of civilizations in the universe is likely wrong, as intelligent life has just begun appearing in our universe. Evolution of advanced organisms has accelerated via development of additional information-processing systems: epigenetic memory, primitive mind, multicellular brain, language, books, computers, and Internet. As a result the doubling time of human functional complexity has reached ca. 20 years. Finally, we discuss the issue of the predicted "technological singularity" and give a biosemiotics perspective on the increase of life's complexity.

## 1. The increase of genetic complexity follows Moore's law

Biological evolution is traditionally studied in two aspects. First, paleontological records show astonishing changes in the composition of major taxonomic groups of animals and plants deposited in sedimentary rocks of various ages (Valentine, 2004; Cowen, 2009). Aquatic life forms give rise to the first terrestrial plants and animals, amphibians lead to reptiles including dinosaurs, ferns lead to gymnosperms and then to flowering plants. Extinction of dinosaurs is followed by the spread of mammals and flying descendants of dinosaurs called birds. Second, Darwin's theory augmented with statistical genetics demonstrated that heritable changes may accumulate in populations and result in replacement of gene variants (Mayr, 2002). This process drives microevolution, which helps species to improve their functions and adjust to changing environments. But despite the importance of these two aspects of evolution, they do not capture the core of the macroevolutionary process, which is the increase of functional complexity of organisms.

Function can be defined as a reproducible sequence of actions of organisms that satisfies specific needs or helps to achieve vital goals (e.g., capturing a resource or reproduction) (Sharov, 2010). To be passed on from one generation to the next, functions have to be encoded within the genome or other information carriers. The genome plays the role of intergeneration memory that ensures the preservation of various functions. Other components of the cell (e.g., stable chromatin modifications, gene imprinting, and assembly of the outer membrane (Frankel, 1989; (Grimes and Aufderheide, 1991)) may also contribute to the intergeneration memory, however, their informational role is minor compared to the genome for most organisms. Considering that the increase of functional complexity is the major trend in macroevolution, which seems applicable to all kinds of organisms from bacteria to mammals, it can be used as a generic scale to measure the level of organization. Because functions are transferred to new generations in the form of genetic memory, it makes sense to consider genetic complexity as a reasonable representation of the functional complexity of organisms (Sharov, 2006; (Luo, 2009). The mechanism by which the genome becomes more complex probably relies heavily on duplication of portions of DNA ranging from parts of genes to gene cascades to polyploidy (Ohno, 1970; (Gordon, 1999), followed by divergence of function of the copies. Developmental plasticity and subsequent genetic assimilation also play a role (West-Eberhard, 2002).

We then have to ask what might be a suitable parameter, measurable from a genome, that reflects its functional complexity? Early studies of the genomes of various organisms showed little correlation between genome length and the level of organization. For example, the total amount of DNA in some single-cell organisms is several orders of magnitude greater than in human cells, a phenomenon known as C-value paradox (Patrushev and Minkevich, 2008). Sequencing of full genomes of eukaryotic organisms showed that the total amount of DNA per cell is not a good measure of information encoded by the genome. The genome includes numerous repetitive elements (e.g., LINE, LTR, and SINE transposones), which have no direct cellular functions; also some portions of the genome may be represented by multiple copies. Large single-cell organisms (e.g., amoeba) need multiple copies of the same chromosome to produce the necessary amount of mRNA. In eukaryotes, DNA has additional functions besides carrying genes and regulating their expression. These non-informational functions include structural support of nuclear matrix and nuclear lamina, chromosome condensation, regulation of cell division and homologous recombination, maintenance and regulation of telomeres and centromeres (Cavalier-Smith, 2005; (Rollins, *et al.*, 2006; (Patrushev and Minkevich, 2008). Segments of DNA with non-genetic functions are mostly not conserved and include various transposing elements as well as tandem repeats. While the ENCODE project has uncovered many functions for 80% of the noncoding DNA in humans (Pennisi, 2012), it has not yet addressed the C-value paradox. Thus we stick to the suggestion to measure genetic complexity by the length of functional and non-redundant DNA sequence rather than by total DNA length (Adami, *et al.*, 2000; Sharov, 2006). A correction for the informational value of noncoding DNA will have to wait for future work on ENCODE studies for a spectrum of species.

If we plot genome complexity of major phylogenetic lineages on a logarithmic scale against the time of origin, the points appear to fit well to a straight line (Sharov, 2006) (Fig. 1). This indicates that genome complexity increased exponentially and doubled about every 376 million years. Such a relationship reminds us of the exponential increase of computer complexity known as a "Moore's law" (Moore, 1965; Lundstrom, 2003). But the doubling time in the evolution of computers (18 months) is much shorter than that in the evolution of life.

What is most interesting in this relationship is that it can be extrapolated back to the origin of life. Genome complexity reaches zero, which corresponds to just one base pair, at time ca. 9.7 billion years ago (Fig. 1). A sensitivity analysis gives a range for the extrapolation of ±2.5 billion years (Sharov, 2006). Because the age of Earth is only 4.5 billion years, life could not have originated on Earth even in the most favorable scenario (Fig. 2). Another complexity measure yielded an estimate for the origin of life date about 5 to 6 billion years ago, which is similarly not compatible with the origin of life on Earth (Jørgensen, 2007). Can we take these estimates as an approximate age of life in the universe? Answering this question is not easy because several other problems have to be addressed. First, why the increase of genome complexity follows an exponential law instead of fluctuating erratically? Second, is it reasonable to expect that biological evolution had started from something equivalent in complexity to one nucleotide? And third, if life is older than the Earth and the Solar System, then how can organisms survive interstellar or even intergalactic transfer? These problems as well as consequences of the exponential increase of genome complexity are discussed below.

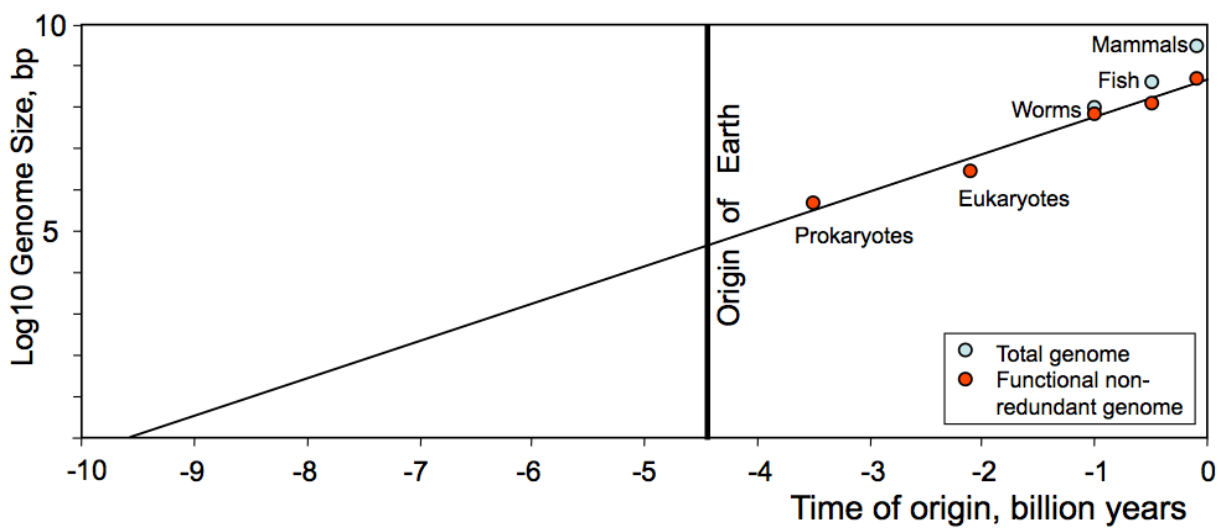

Figure 1. On this semilog plot, the complexity of organisms, as measured by the length of functional non-redundant DNA per genome counted by nucleotide base pairs (bp), increases linearly with time (Sharov, 2012). Time is counted backwards in billions of years before the present (time 0). Modified from Figure 1 in (Sharov, 2006).

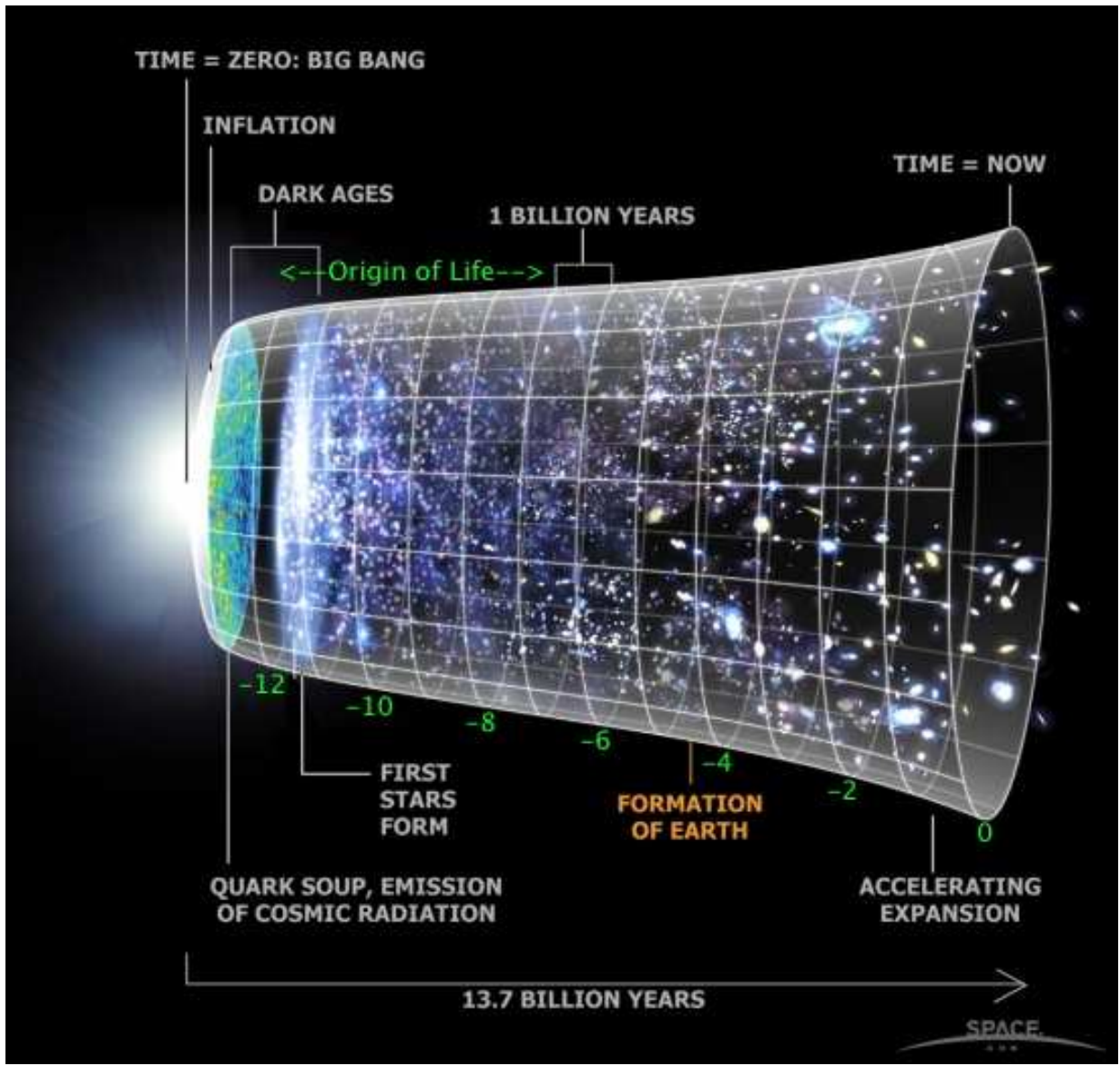

Figure 2. A schematic view of the development of the universe since the Big Bang, courtesy of the Hubble Space Telescope Science Institute, on which we have superimposed our estimate for the origin of life, $9.7 \pm 2.5$ billion years ago. Note that the "Dark Ages" may have ended at -13.55 billion years (Zheng, *et al.*, 2012) (with the Big Bang at -13.75 billion years (Jarosik, *et al.*, 2011)), rather than at -11.5 billion years, as depicted.

## 2. How variable are the rates of evolution?

To extrapolate of the rate of biological evolution into the past, we need to provide arguments why this rate is stable enough. There is no consensus among biologists on the question how variable are the rates of evolution. Darwin thought that in general evolutionary changes accumulate gradually through a series of small steps, rather than by sudden leaps (Darwin, 1866). However, he also pointed out that the rate of evolution is not uniform: "But I must here remark that I do not suppose that the process ever goes on so regularly as is represented in the diagram, though in itself made somewhat irregular, nor that it goes on continuously; it is far more probable that each form remains for long periods unaltered, and then again undergoes modification" (Darwin, 1872). This concept, now called punctuated equilibrium, assumes a high variation in the rates of evolution (Gould and Eldredge, 1977). Paleontological records indicate that major evolutionary changes occurred during very short intervals that separated long epochs of relative stability. If the concept of punctuated equilibrium is applied to the global trend of the increase of functional complexity of organisms (Fig. 1), then it may be argued that rates of evolution are so unstable that any extrapolation of them into the past is meaningless. In particular, it was suggested that the rates of primordial evolution were much higher than normal simply because of the absence of competition (Koonin and Galperin, 2003). The notion of unusually rapid primordial evolution was suggested also by other scientists (Davies, 2003; Lineweaver and Davis, 2003). These attempts to explain the presumed origin of life on Earth are strikingly similar to stretching and shrinking of time scales in Biblical Genesis to fit preconceptions (Schroeder, 1990).

Although we fully agree that evolutionary rates fluctuate in time and that catastrophic changes of the environment followed by mass extinction provide a boost of novel adaptations to survived lineages, we strongly disagree that the concept of punctuated equilibrium is applicable to the general trend of the increase of functional complexity of organisms (Fig. 1). First, adaptive radiation of lineages observed during periods of rapid evolutionary change has nothing to do with the increase of functional complexity. Multicellular organisms have enough functional plasticity to produce a large variety of morphologies based on already existing molecular and cellular mechanisms. Second, many rapid changes in the composition of animal and plant communities resulted from migration and propagation of already existing species (Dawkins, 1986), a mechanism that does not require an increase in functional complexity. Third, there is no reason to expect that functional complexity of organisms did not increase during long "equilibrium" periods with no dramatic change in the morphology of organisms. Morphology is the tip of the evolutionary iceberg as the greatest changes occur at the molecular level. The common idea that stabilizing selection simply preserves the status quo in evolution is based on the misunderstanding of the original theory of stabilizing selection (Schmalhausen, 1949). Stabilizing selection leads to increased plasticity of organisms (West-Eberhard, 2002) which is achieved via novel signaling pathways that replace less reliable old pathways. All these changes may have no immediate effect on morphology, but nevertheless these are real changes that lead to the increase of functional complexity. Schmalhausen developed his theory without knowledge of molecular biology, which was not available at that time. But he managed to capture the idea on how phenotypic plasticity reshaped evolution. Finally, there is a difference in time scales: punctuated equilibrium refers to relatively short periods of evolutionary change (millions of years), whereas the global growth of functional complexity becomes apparent at the time scale of billions of years.

The reason why living organisms cannot increase their functional complexity instantly may be that it takes a long time to develop each new function via trial and error. Thus, simultaneous and fast emergence of numerous new functions is very unlikely. In particular, the origin of life was then not a single lucky event but a gradual increase of functional complexity in evolving primordial systems. Similarly, the emergence from prokaryotes to eukaryotes was not the result of one successful symbiosis, but may have involved as many as 100 discrete innovative steps (Cavalier-Smith, 2010). This view is consistent with Darwin's insight that early evolution was slow and gradual:

> "During early periods of the earth's history, when the forms of life were probably fewer and simpler, the rate of change was probably slower; and at the first dawn of life, when very few forms of the simplest structure existed, the rate of change may have been slow in an extreme degree. The whole history of the world, as at present known, although of a length quite incomprehensible by us, will hereafter be recognised as a mere fragment of time, compared with the ages which have elapsed since the first creature, the progenitor of innumerable extinct and living descendants, was created" (Darwin, 1866).

Another factor that may have reduced the rates of primordial evolution was the absence of well-tuned molecular mechanisms, which are now present in every cell. In particular, there was no basic metabolism to produce a large set of simple organic molecules (e.g., sugars, amino acids, nucleic bases), and no template-based replication of polymers (see section 4). These two obstacles substantially reduced the frequency of successful "mutations" and, as a result, the initial rate of complexity increase was likely even slower than shown in Fig. 1. Thus, there is no basis for the hypothesis that the evolution of such complex organisms as bacteria with genome size of ca. $5 \times 10^5$ bp could have been squeezed into <500 million years after Earth's cooling.

**3. Why did genome complexity increase exponentially?**

The increase of functional complexity in evolution can be modeled on the basis of known mechanisms, which appear to act as positive feedbacks (Sharov, 2006). First, the model of a hypercycle considers a genome as a community of mutually beneficial (i.e., cross-catalytic) self-replicating elements (Eigen and Schuster, 1979). For example, a mutated gene that improves proofreading of the DNA increases the replication accuracy not only of itself but also of all other genes. Moreover, these benefits are applied to genes that may appear in the future. Thus, already existing genes can help new genes to become established, and as a result, bigger genomes grow faster than small ones. Second, new genes usually originate via duplication and recombination of already existing genes in the genome (Ohno, 1970; Patthy, 1999; Massingham, *et al.*, 2001). Thus, larger genomes provide more diverse initial material for the emergence of new genes. Third, large genomes support more diverse metabolic networks and morphological elements (at various scales from cell components to tissues and organs) than small genomes, which in turn, may provide new functional niches for novel genes. For example, genes in multicellular organisms operate in highly diverse environments represented by various types of cells and tissues. Progressive differentiation of cells supports the emergence of gene variants that either perform the same function in specific cell types or modify the original function for specific needs of some cells. Replication followed by divergence of differentiation trees allows duplication of whole cell types, followed by them assuming different functions within an organism (Gordon, 1999; (Gordon and Gordon, 2013). These mechanisms of positive feedback may be sufficient to cause an exponential growth in the size of functional nonredundant genome.

Existing data also indicates that the genetic complexity may have increased a little faster than exponentially (i.e., hyperexponentially), which may be explained by phase transitions to higher levels of functional organization (Sharov, 2006; Markov*, et al.*, 2010). For example, the time of genome doubling in Archae and Eubacteria was 1080 and 756 million years, respectively (these estimates are based on the largest known archaeal genome, 5 Mb, in *Methanogenium frigidum* and bacterial genome, 13 Mb, in *Sorangium cellulosum*) (Bernal*, et al.*, 2001). These estimates are 2.9 and 2.0 fold longer than the doubling time in Eukaryota. The difference between the rates of increase of genome complexity between most successful and lagging lineages can be explained by evolutionary constraints of the latter ones (e.g., inefficient DNA proofreading and absence of mitosis). Thus, the rate of the "complexity clock" may have increased with the emergence of eukaryotes, and therefore, life may have originated even earlier than expected from the regression in Fig.1. That would push the projected origin of life close to the origin of our galaxy and the universe itself. Thus, life may have originated shortly after parts of the universe cooled down from the Big Bang (Gordon and Hoover, 2007). For the sake of this chapter, we are assuming that the Big Bang model for the universe and its age of $13.75 \pm 0.11$ billion years is correct (Jarosik*, et al.*, 2011), although some evidence suggests that our universe is substantially older (Kazan, 2010).

The exponential increase of functional complexity is consistent with Reid's view of evolution as cascading emergences:

> "As evolution progresses, the freedom of choice increases exponentially…. intrinsic complexification of differentiated cell types, is overall an exponential function of reproduction and time-quite a simple equation…. the historical curve of some lineages, especially that of hominids, fits the simple exponential equation, its logarithmic slope theoretically determined by the fact that the acceleration of complexification is virtually equivalent to increasing adaptability and freedom to explore unexploited environments" (Reid, 2007).

## 4. Could life have started from the equivalent of one nucleotide?

Autocatalytic synthesis (in contrast to decay) is a rare property among organic molecules. Thus, it was suggested that it can arise more easily in multi-component mixtures of molecules with random cross-catalysis (Kauffman, 1986; Kauffman*, et al.*, 1986). In particular, Kauffman suggested that a mixture of peptides could form a closed autocatalytic set, where the synthesis of each component is catalyzed by some members of the same set. Members of autocatalytic sets are not necessarily peptides; they can be RNA oligonucleotides (Lincoln and Joyce, 2009) or any other kind of organic molecules.

But despite the attractiveness of the idea that life originated from autocatalytic sets and elaborate mathematical support (Mossel and Steel, 2007), there are serious problems with this hypothesis. First, we focus on the most common version of this hypothesis where elements of autocatalytic sets are heteropolymers (e.g., peptides or oligonucleotides). The "RNA world" hypothesis assumes that first living systems had self-replicating nucleic acids (Gilbert, 1986) or other kinds of similar heteropolymers (TNA, PNA) (Nelson, *et al.*, 2000; Orgel, 2000). Although some RNA molecules can catalyze the polymerization of other RNA (Johnston, *et al.*, 2001), this reaction requires abundant free nucleotides. Nucleotides can be synthesized abiogenically (Powner, *et al.*, 2009), but they are unlikely to become concentrated in quantities sufficient to support RNA polymerization in a population of proto-organisms. Even if several molecules appear in close proximity to each other due to a once-in-a-universe lucky coincidence and produce a complimentary RNA strain, there would be no nucleotides left to make the next generation of replicons. Nucleotides can be synthesized from bases and sugars by RNA-mediated catalysis (Unrau and Bartel, 1998), but both bases and sugars are rare molecules which are unlikely to be supplied in sufficient quantities. Polymers like nucleic acids and peptides may persist only on condition of an unlimited supply of monomers, and this requires a heritable mechanism for their synthesis from simple and abundant organic and non-organic resources (Copley, *et al.*, 2007; Sharov, 2009). Thus, the emergence of polymers was the second chapter in the history of life, whereas the first chapter was the origin of simple molecules that supported both metabolic and hereditary functions (Jablonka and Szathmáry, 1995; Sharov, 2009).

The next question is whether self-replicating autocatalytic sets can become assembled from simple molecules. Alas, most abundant organic molecules in the non-living world (e.g., saturated hydrocarbons) are inert. Catalytically active organic molecules are rare and often unstable, thus it is unlikely that they would become concentrated together in a tiny space to become integrated into an autocatalytic set. High concentrations and enclosures are needed to increase the rates of mutual catalysis so that they compensate for the loss of molecules due to their degradation and diffusion.

A more recent "GARD" (Graded Autocatalysis Replication Domain) model of the origin of life is based on "compositional assemblies" of simple lipid-like molecules (Segré and Lancet, 1999; Segré, *et al.*, 2001; Bar-Even, *et al.*, 2004). In contrast to autocatalytic sets that exist in a homogeneous space. the GARD model assumes a phase separation between aggregates of molecules (e.g., lipid microspheres) and a homogeneous environment. The GARD model is substantially more realistic than autocatalytic sets because the requirement of catalytic closure is replaced by the assumption of selective attraction or repulsion of molecules to or from the aggregates (i.e., partitioning), which is a more widespread behavior in simple molecules. Models of growth of compositional assemblies usually also assume that large assemblies can break down into two (or more) daughter assemblies. As a result, these assemblies can reproduce and form discrete quasispecies. Disproportional split of components between daughter systems can be viewed as "mutations" that may occasionally give rise to new quasispecies. However, it appears that compositional assemblies lack evolvability as the number of potential quasispecies is limited (Vasas, *et al.*, 2010). From the systems point of view, the growth of compositional assemblies is similar to the growth of crystals. The same chemical may produce different crystals (quasispecies), local defects in a crystal lattice may give rise to a new crystal type (Cairns-Smith, 1982). Similar to compositional assemblies, crystals lack evolutionary potential.

The evolution of primordial living systems requires heredity, but neither nucleic acids nor other complex polymers were initially available to support it. Thus, hereditary functions must have been carried out by simpler molecules. Heredity requires autonomous self-production, which is a generalization of autocatalytic synthesis. In chemical terms, compositional assemblies in the GARD model have no catalysis, but in systems terms, there is an autocatalysis of assemblies as they produce assemblies with matching composition.

The general notion of self-reproduction has been defined using the formalism of Petri nets (Sharov, 1991). In short, a system is self-reproducing if there is a finite sequence of transitions (i.e., reactions) that results in the increase of the numbers of all components within the system. For example, the formose reaction is autocatalytic and makes sugars from formaldehyde (Huskey and Epstein, 1989). Such reactions can propagate in space, which is similar to the growth and expansion of populations of living organisms (Gray and Scott, 1994; Tilman and Kareiva, 1997). Autocatalytic reactions have two alternative steady states: "on" and "off" (the "on" state is stabilized via a limited supply of resources). Thus, they represent the most simple hereditary system or memory unit (Jablonka and Szathmáry, 1995; Lisman and Fallon, 1999). For example, a reverse citric acid cycle, which captures carbon dioxide and converts it into sugars, may become self-sustainable, at least theoretically (Morowitz, *et al.*, 2000). Prions are examples of autocatalytic reproduction (Griffith, 1967; Laurent, 1997; Watzky, *et al.*, 2008), and indeed have been invoked in various ways in speculations on the origin of life (Steele and Baross, 2006; Maury, 2009; Hu, *et al.*, 2010). However, they cannot support the synthesis of the primary (i.e., unfolded) polypeptide.

Autocatalysis is necessary for the origin of life, but not sufficient. The specific feature of autocatalysis in living systems is that it is linked functionally with a local environment (e.g., cell), and this linkage can be viewed as a coding relation (Sharov, 2009). In particular, the autocatalytic system modifies (encodes) its local environment, and this modification increases the rate of autocatalysis. This functional linkage is a necessary condition for cooperation between multiple autocatalytic components if they happen to share their local environment. In economic terms, the system invests in the modification of its environment, and therefore cannot leave its investment. This can also be viewed as a "property" relation at the molecular level. The autocatalytic system is the owner of its local environment, which plays the role of "home" or "body". Because the system is attached to its home, it is forced to cooperate with other autocatalytic systems that may appear in the same local environment. The local environment can be represented by either enclosure or attachment to a surface. Although all known free-living organisms have enclosures (cell membranes), life may have started from surface metabolism because autocatalysis has a much higher rate on a two-dimensional surface than in three-dimensional space (Wächtershäuser, 1988), an example of dimension reduction (Adam and Delbrück, 1968).

We have proposed a "coenzyme world" scenario where hereditary functions are carried out by autocatalytic molecules, named "coenzyme-like molecules" (CLMs) since they are catalytically active and may resemble existing coenzymes (Sharov, 2009). Because many coenzymes (e.g.,

ATP, NADH, and CoA) are similar to nucleotides, CLMs can be viewed as predecessors of nucleotides. The most likely environments for CLMs were oil (hydrocarbon) microspheres in water because (1) hydrocarbons are the most abundant organic molecules in the universe (Deamer, 2011) and are expected to exist on early terrestrial planets (Marcano, *et al.*, 2003), (2) oil microspheres self-assemble in water and (3) it is logical to project the evolutionary transformation of oil microsphere into a lipid membrane. CLMs can colonize the surface of oil microspheres in water as follows. Assume that rare water-soluble CLMs cannot anchor to the hydrophobic oil surface. However, some microspheres may include a few fatty acids with hydrophilic ends that allow the attachment of CLMs (Fig. 3). Once attached, a CLM can catalyze the oxidation of outer ends of hydrocarbons in the oil microsphere, thus providing the substrate for binding of additional CLMs. Accumulation of fatty acids increases the chance of a microsphere to split into smaller ones, and small microspheres can infect other oil microspheres, i.e., capture new oil resource. This process of autocatalytic adhesion creates a two-level hierarchical system, where CLMs play the role of coding elements. Alternatively, CLMs can be synthesized from precursors (e.g., from two simpler molecules A and B) on the surface of microspheres. For example, when molecule A becomes attached to the surface of a microsphere, it changes conformation so that it can interact with another water-soluble molecule B. As a result, the synthesis of A + B => AB is catalyzed by the oil microsphere. If the product AB is capable of oxidizing hydrocarbons into fatty acids, then the whole system becomes autocatalytic.

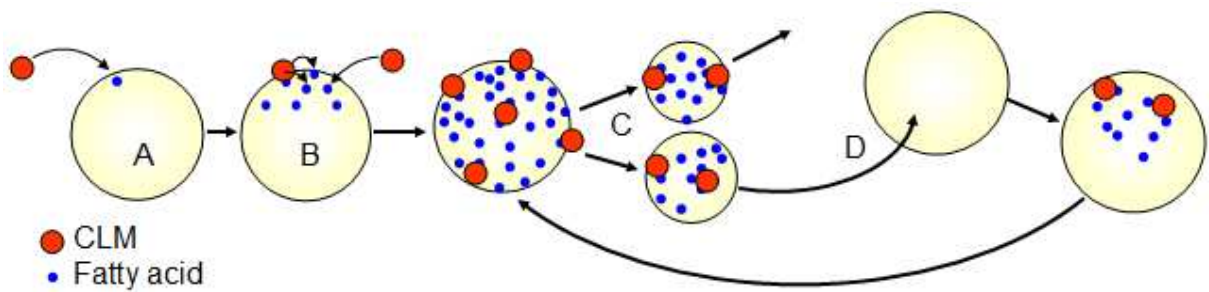

Figure 3. Coenzyme world: coenzyme like molecules (CLMs) on the oil microsphere. (A) CLM can anchor to the oil microsphere via rare fatty acid molecules. (B) The function of a CLM is to oxidize hydrocarbons to fatty acids, which provides additional anchoring sites for new CLMs. (C) Accumulation of fatty acids increases the chance of a microsphere to split into smaller ones, and (D) small microspheres can infect other oil microspheres (i.e., capture new oil resource).

Several kinds of autocatalytic coding elements may coexist on the same oil microsphere, creating a system with combinatorial heredity (Sharov, 2009). Each kind of coding element performs a specific function (e.g., capturing resource, storing energy, or catalysis of a reaction) and ensures the persistence of this function. However, coding elements are not connected, and hence, are transferred to offspring systems in different combinations. Despite random transfer, the combinatorial heredity can be stable because (1) coding elements are present in multiple copies and therefore each offspring has a high probability of getting the full set, and (2) natural selection preserves preferentially systems with a full set of coding elements. The efficiency of the latter mechanism was shown in a "stochastic corrector model" (Szathmáry, 1999). New types of coding elements can be added by (1) acquisition of entirely new CLMs from the environment, (2) modification of existing CLMs and (3) polymerization of CLMs. Combinatorial heredity can eventually lead to the emergence of synthetic polymers (Sharov, 2009). For example, if a new CLM, C, can catalyze the polymerization of another CLM, A, then together they encode long polymers AAAAA..., which can cover the surface of the microsphere and substantially modify its physical properties.

**5. How heritable surface metabolism may have evolved into RNA-world cell?**

The surfaces of oil microspheres provide discrete and abundant local environments for emerging primordial life. However, further evolution of life is constrained by (1) limited amount of abiotic liquid hydrocarbons, and (2) limited dimensionality (i.e., 2D) of the functional system. Limited abundance of hydrocarbons prevented primordial organisms from increasing in size, which is a serious constraint on the surface area and, consequently, the total rate of metabolism. Although surface metabolism is beneficial for the emergence of primordial life, it cannot support the storage of a large amount of resources. Also, the movement of polymers on the surface is highly restricted because two long molecules adsorbed at points along their lengths cannot easily pass each other in a 2D-space.

Both problems could have been solved with one invention: a transition from surface metabolism to a membrane-enclosed cell. From the topological point of view, an oil microsphere can be converted into a membrane-enclosed cell via engulfing water. Such "double" microspheres are easily generated by agitating an emulsion of liquid hydrocarbons in water but they are not stable. Cell-like organization yields little functional advantage if the membrane breaks before the cell divides. Thus, the membrane has to be strong enough to sustain mechanical disturbances, and the volume of a cell has to be controlled to prevent bursting. Hypothetically, primordial systems passed through a transitional period with relatively unstable membranes, but managed to secure short-term benefits from temporary cell-like organization (e.g., via storage of accumulated energy at the membrane). During this period, the major selection pressure was to develop more stable membranes, which required the synthesis of glycerol, which is the backbone of lipids, and phospholipids. Emergence of a heritable metabolism that generates glycerol was the major evolutionary achievement at that time. Later in evolution, glycerol could have been used to make sugars, which are suitable for storing energy and regulating osmosis.

Initial steps of primordial evolution were slow and inefficient because there was no universal rule for producing new coding elements. Some improvement was likely achieved by transformation of old coding elements into novel ones via modification of functional groups or polymerization (Sharov, 2009). However, there was no streamlined procedure for making new coding elements until the invention of template-based (or digital) replication. Replication is a special case of autocatalysis, where each coding element is a linear sequence of a few kinds of monomers, and copying is done sequentially via predefined actions applied to each monomer (Szathmáry, 1999). Digital replication makes the coding system universal because it works for any sequence. Thus, there is no need to invent individual recipes for copying modified coding molecules.

The starting point for the origin of template-based replication was the existence of polymeric coding elements with either random or repetitive sequence (Sharov, 2009). Polymers may initially stick to each other to perform some other functions (e.g., to increase the stability of the membrane or facilitate polymerization). The shorter strand of the paired sequence can then become elongated by adding monomers that weakly match to the overhanging longer strand. Then the selection supported the increase in specificity of this process and helped to produce better copies of existing polymers. Invention of digital replication, therefore, may have been the turning point in the origin of life that substantially increased the hereditary potential of primordial living systems (Jablonka and Szathmáry, 1995; Sharov, 2009). It is not clear if template-based replication of polymers appeared before or after the formation of a stable cell membrane. But in any case, these two evolutionary events produced cells with ribozyme-based catalysts, and this stage of evolution is known as "RNA-world" (Gilbert, 1986). These organisms still had no tools for protein synthesis, but they likely were able to make simple peptides with the help of ribozymes.

The Last Universal Common Ancestor (LUCA) of all life forms on Earth (Doolittle, 2000; Theobald, 2010) is far more complex than RNA-world organisms described above. Its heredity is based on DNA rather than RNA, and it has fully developed protein synthesis programmed by genes. The integrity of the DNA is maintained by a group of "maintenance" enzymes, including DNA topoisomerase, DNA ligase and other DNA repair enzymes. The RNA-polymerase complex is used to synthesize mRNA copies of each gene, and then each mRNA is translated into a protein by the action of ribosomes and tRNA. LUCA can make fatty acids from sugars and therefore it is no longer dependent on the supply of abiotically synthesized hydrocarbons. The outer membrane of LUCA includes ion pumps. Although the genome of LUCA is a product of phylogenetic reconstruction, it is possible that bacterial spores that arrived to the primordial Earth from cosmic sources had functional complexity similar to LUCA. Based on the rate of complexity increase (Fig. 1), we expect that RNA-world cells emerged ca. 2 billion years after the origin of life, and LUCA-like organisms appeared ca. 3 billion years later.

**6. How can organisms survive interstellar transfer?**

Bacterial spores have unusually high survival rates even in very harsh conditions, and therefore, they are most likely candidates for interstellar transfer. Contaminated material can be ejected into space from a planet via collision with comets or asteroids (Ehrlich and Newman, 2008). Then bacterial spores may remain alive in a deep frozen state for a long time that may be sufficient for interstellar transfer. Bacterial spores were reported revived after 25-35 million years of dormancy (Lambert, *et al.*, 1998). A more recent discovery of viable bacteria trapped in the 0.75 million year old ice (Katz, 2012) suggests that bacteria may be preserved in ice for long times. One of the scenarios of life's arrival to new planets is the capture of small contaminated particles by a protoplanetary disc before planet formation (Wallis and Wickramasinghe, 2004).

The hypothesis of panspermia becomes more plausible if the Solar System originated from the remnants of the exploded parental star (Joseph, 2009). Remnants of planets from exploded supernovae (Gordon and Hoover, 2007) can carry billions of bacterial spores and maybe even active chemosynthetic bacteria deep beneath the surface. Metabolically active bacteria are able to repair their DNA and withstand the damaging effect of natural soil radiation (Johnson*, et al.*, 2007). If the Earth incorporated some of these planet fragments it could have been seeded by a diverse community of bacterial species and their viruses. Prediction and discovery of rogue planets (Gibson, 2000; Samuel, 2001; Debes and Sigurdsson, 2002; Gladman and Chan, 2006; Vanhamaki, 2011) that could harbor life (Abbot and Switzer, 2011; Badescu, 2011) strengthen the panspermia hypothesis which can no longer be dismissed on the basis of disbelief (McNichol and Gordon, 2012).

**7. Implications of the cosmic origin of life on Earth**

Contamination with bacterial spores from space appears the most plausible hypothesis that explains the early appearance of life on Earth. Thus, despite the fact that we don't have a final answer, it makes sense to explore the implications of a cosmic origin of life, before the Earth existed. First, we conclude that life took a long time, perhaps 5 billion years, to reach the complexity of bacteria. Thus, the possibility of repeated and independent origins of life of this complexity on other planets in our Solar System can be ruled out. Extrasolar life is likely to be present at least on some planets or satellites within our Solar System, because (1) all planets had comparable chances of being contaminated with microbial life, and (2) some planets and satellites (e.g., Mars, Europa, and Enceladus) provide niches where certain bacteria may survive and reproduce. If extraterrestrial life is present in the Solar System, it should have strong similarities to terrestrial microbes, which is a testable hypothesis. We expect that they have the same nucleic acids (DNA and RNA) and similar mechanisms of transcription and translation as in terrestrial bacteria. The ability to survive interstellar transfer was the major selection factor among prokaryotes on the cosmic scale. Thus, bacterial life forms were successful in colonizing the cosmos only if they were resilient to radiation, cold, drying, toxic substances, and highly adaptable to a broad range of planetary environments. In particular, photosynthesis or chemosynthesis is needed to be independent from organic resources. The similarity between terrestrial and extraterrestrial bacteria may appear sufficient to draw a unified evolutionary tree of life, though it may be complicated by later transfers between the planets, such as between Earth and Mars (Gordon and McNichol, 2012).

Second, there was no intelligent life in our universe at the time of the origin of Earth, because the universe was 8 billion years old at that time, whereas the development of intelligent life requires ca. 10 billion years of evolution. Thus, the idea that life was transferred to Earth by intelligent beings (i.e., "directed panspermia") (Crick and Orgel, 1973) is incorrect.

Third, attempts to reproduce the origin of life in laboratory conditions (Damer*, et al.*, 2012) may prove more difficult than it is generally expected because such experiments have to emulate many cumulative rare events that occurred during several billion years before organisms reached the complexity of the RNA world (Sharov, 2009), contradicting the notion that life evolved rapidly on Earth (Davies, 2003; Lineweaver and Davis, 2003). Despite the success of copying an existing bacterial genome (Gibson*, et al.*, 2008), humans have so far failed to invent a single new functional enzyme from scratch (i.e., without copying it from nature), and have had limited success in imitating existing enzymes (Bjerre*, et al.*, 2008). Thus, it would be hard to make a primitive living system that does not resemble anything that we observe on Earth. A more realistic approach is to use chemical reactors, where primordial systems with heritable functions may emerge spontaneously from abiogenic components. These reactors may require spatial and temporal heterogeneity and dynamic interaction between multiple liquid and solid components. Special sensors should be designed to monitor the composition of molecular assemblies to detect irreversible and sustainable changes.

Fourth, the original Drake equation for guesstimating the number of civilizations in our galaxy (Wikipedia contributors, 2012c) may be wrong, as we conclude that intelligent life like us has just begun appearing in our universe. The Drake equation is a steady state model, and we may be at the beginning of a pulse of civilization. Emergence of civilizations is a non-ergodic process, and some parameters of the equation are therefore time-dependent. Because the cosmic transport of life is most likely limited to prokaryotes, young planets have not had enough time to develop intelligent life. Another time-dependent process is the probability of interstellar transfer of bacteria, which we expect to have become more frequent as the total pool of bacteria in the galaxy increased with time. There are many modifications of the Drake equation, but if civilizations have just begun to appear, any version is of limited use. The answer to the Fermi paradox (Wikipedia contributors, 2012d) may be that we are amongst the first, if not the only so far, civilization to emerge in our galaxy. The "Rare Earth" hypothesis (Ward and Brownlee, 2003) need not be invoked. The linking of civilization to the lifetime of a particular star, such as our Sun (Livio and Kopelman, 1990; Webb, 2002), is also not necessary.

Fifth, the environments in which life originated and evolved to the prokaryote stage may have been quite different from those envisaged on Earth. Thus, emulating conditions on the young Earth may not increase the chance of generating primordial living systems in the lab. Even a bigger mistake would be to use contemporary minerals in such experiments because most mineral species on Earth are directly or indirectly modified by life (Hazen, 2010). To define possible environments for the origin of life we can extrapolate the evolution of minerals back in time to the initial cooling of the universe after the Big Bang. The major questions to ask are: (1) When did stars and planets form? (2) What was the elemental composition of stars and planets versus cosmic time? (3) How was the surface of those early planets stratified? (4) What atmospheres might those planets have had? It is reasonable to assume that life originated in the presence of water, as water is very abundant in space and is a byproduct of star formation. Young stars shoot jets of water into the interstellar space (Fazekas, 2011). Large quantities of water have been detected in space clouds (Glanz, 1998). Thus, it is reasonable to assume that water was present early in the young universe and could have supported the origin of life. Major chemical elements of living organisms (carbon, hydrogen, oxygen, and nitrogen) are among the most abundant in the Universe. Phosphorus is less abundant and may have been the limiting factor for the origin of life, although recent studies suggest that it can be effectively extracted from apatite or come from volcanic activity or space (Schwartz, 2006). Life may have started on planets around the first, low "metal" stars, where "metals" mean, to astronomers, higher atomic number elements that formed from hydrogen and helium during stellar nucleosynthesis. Such stars may have formed as early as 200 million years after the Big Bang (Zheng, *et al.*, 2012). High metallicity is a negative factor for life origin because Earth-like planets may be destroyed by giant planets (Linewater, 2001). The stars of such planets probably lasted only 4 billion years, so if life didn't have a "false start" in such systems (Johnson and Li, 2012), it may have been propagated into interstellar space during the star's supernova event (Gordon and Hoover, 2007). Therefore, the time scale we are proposing for the origin and complexifying of life requires panspermia mechanisms (McNichol and Gordon, 2012) for life to persist.

## 8. Genetic complexity lags behind the functional complexity of mind

The idea that genetic complexity can be used as a generic scale of functional complexity of all organisms is intellectually attractive and works well at the lower end of complexity. However, the genome fails to capture correctly the complexity of higher level organisms. Based on Fig. 1, the difference in genetic complexity is small between fish and mammals, and there is no difference between mouse and human. What makes humans superior to mice is not the genome but mind.

Eukaryotes progressively used epigenetic memory (i.e., stable chromatin modifications) to encode or modify their functions. In contrast to the genome, the epigenetic memory is rewritable, and therefore can easily support phenotypic plasticity and development of habits. Eventually, epigenetic mechanisms led to the emergence of mind which is a tool for classifying and modeling of objects (Sharov, 2012). Mind operates at the level of holistically perceived objects, whereas primitive organisms (e.g., bacteria) regulate their functions via simple molecular level signal-response circuits. The power of mind has increased dramatically since the emergence of higher-level learning which enabled organisms to distinguish and model new classes of objects. As a result, the functional complexity of organisms became encoded partially in the heritable genome and partially in the perishable mind. Despite the short life span of the individual mind, its advantages are tremendous because it allows an organism to develop complex behaviors such as running, flying, and social interactions, and helps to adjust these behaviors to changing environments. As the complexity of mind increased, the role of the genome has shifted from direct coding and controlling of functions to creating favorable conditions for the development of mind, which can take care of these functions later in life. In other words, the role of the genome became to provide the informational infrastructure (e.g, nervous system) and initial training for the growing mind. As a result, the functional complexity of higher animals (e.g., birds and mammals) started growing much faster than in their ancestors, and this growth can no longer be captured by the growth of genetic complexity (Fig. 1). This is an example of accelerating (or hyperexponential) growth of complexity caused by the emergence of novel methods of information processing.

The increase of the functional complexity of mind is more difficult to measure than the increase of genetic complexity because we still do not know how mental memory is encoded. The size of the brain can be used as a first approximation of the complexity of mind. However, the size of brain is also correlated to body size (McHenry, 1975), which makes comparisons difficult. Plots of brain mass versus body mass, nevertheless, lead to clear classifications and trends (Jerison, 1973). The functionality of the brain may depend more strongly on its structure (e.g., on surface area, neuron density, or connectedness) than on the volume (Roth and Dicke, 2005). More reliable trends in brain size can be detected within a narrow taxonomic group. For example, the regression of log-transformed brain size to the time of origin in Hominids (data from (Wikipedia contributors, 2012a)) indicates the doubling time of 3.2 million years. If we use encephalization quotient, which adjusts brain volume to body size (Roth and Dicke, 2005), then the doubling time from chimps to humans equals 3.0 million years. Thus, the rate of brain increase in evolution exceeds the rate of genetic evolution by a factor of ca. 100, which shows the advantage of switching from the genome code to the "mind code".

The origin of humans marks another major transition in the evolution of functional complexity because humans invented methods to transfer information effectively between minds. Initially, information transfer was based on copying the behavior of other members of a social group, but later it was augmented by the development of speech and finally by the written language. As a result, the content of minds became shared between individuals and preserved for future generations. This transition fueled further acceleration for the growth of functional complexity. To get an idea of the doubling time of human social information, we consider the number of characters in the Chinese language, which increased from ca. 2,500 at 3200 years ago (the Oracle bone script) to ca. 47,000 at the present time (New World Encyclopedia contributors, 2008). Thus, the rate of language doubling time appears to be 825 years, which exceeds the rate of brain increase in evolution by a factor of >3000.

**9. Extrapolating the growth of complexity into the future**

Predicting the future was historically based on spiritual revelations or dreams (e.g., Joseph, Isaiah, John the Baptist), or analysis of ancient texts (Nostradamus). Nowadays we can use science and statistics to extrapolate existing trends into the future, at least over the persistence time of our models (Pilkey and Pilkey-Jarvis, 2007). For example, Moore noted that the number of components in integrated electronic circuits had doubled every year from 1958 to 1965 (Moore, 1965). Similar exponential trends are known for other technologies (e.g., speed of DNA sequencing, hard disc capacity, and bandwidth of networks). High rates of exponential growth of these technologies with doubling time of ca. one year has led to the idea of a "technological singularity", which refers to the time when technology-based intelligence will emerge and possibly replace humans (Kurzweil, 2005).

However, growth rates of specific technologies should not be confused with the increase of functional complexity of the human civilization as a whole. Humans developed economy mechanisms (e.g., loans, bonds, stock market) that help us to redistribute resources to key industries that are perceived as bottlenecks to functional growth. Recent such industries include computer technology, Internet, wireless communication, and biotechnology. But we cannot expect that these industries will remain major limiting factors forever. For example, computation is no longer the key factor in most human applications; instead it became an inexpensive commodity, which can be easily expanded on demand using cloud computing. Thus, the further progress in computer technology will not be as revolutionary as it was in the previous 3 decades. Similarly, the speed of DNA sequencing will soon reach its limits and cease to be the most critical factor. The rate of increase of functional complexity of human civilization can be better measured by indicators that are not linked to a single technology. For example, the doubling time of the number of scientific publications from 1900 to 1960 was only 15 years (de Solla Price, 1971). Interestingly, extrapolating the exponential increase of scientific publications backwards gives us an estimated origin of science at 1710 which is the time of Isaac Newton. The increase in the number of patents (data from http://en.wikipedia.org/wiki/Patent) has the doubling time of ca. 25 years. Thus, the functional complexity of the human civilizations doubles approximately every generation (i.e., 15-25 years), which is ca. 20 fold slower than for most "critical" technologies.

Prediction of future events such as "technological singularity" (Kurzweil, 2005) is flawed if based on trends within a single technology (in this case, computer speed). There is no doubt that computers can outperform humans at specific tasks ranging from simple number crunching to automatic vehicle navigation and games (e.g., chess). It is also obvious that many human jobs will shift to fully-automated devices, continuing a trend that started in the age of steam (Carnegie, 1905). But this does not mean that technology is going to replace humans because novel jobs are created to operate, program, and use each new technology. Technology cannot compete with humans because humans control the resources and would not allow technical agents to take over. Until technical devices become fully self-reproducing, which is not in the cards so far, they pose no danger for humanity. The only self-reproducing artificial agents so far are computer viruses, and they indeed can do substantial harm within information-processing networks and the equipment they control (Wikipedia contributors, 2012e). But computer viruses cannot displace humans, and most of their effects can be controlled.

Another interesting trend is that human intelligence can be augmented by technology, which includes molecular bioengineering and development of brain-computer interfaces (BCIs). In particular, genetic manipulation and the use of growth factors and hormones during brain development may artificially increase the number of neurons and enhance their connectivity (Chelen, 2012). In this way, an "organic singularity" may be achieved before the predicted "technological singularity". Mixed organic-technical systems based on BCIs also have promising prospects. Animals with implanted electrodes in their brain can learn to manipulate a robotic arm or computer screen (Carmena, *et al.*, 2003). Non-invasive BCI (EEG-based) allows humans to manipulate a computer (Farwell and Donchin, 1988). Functional MRI technology allows instantaneous reconstruction of videos watched by human subjects (Nishimoto, *et al.*, 2011), a form of mind reading. The immediate application for BCI is to compensate for various disabilities (e.g., loss of vision, hearing, movement), but in the future it may serve to augment the intelligence of normal people. In particular, it is attractive to use BCI to create additional vision fields that represent a computer screen or improve deteriorating memory with the help of external silicon memory chips. Further advance in the quality of BCI can be expected if the interface is established early in life, so that brain functions can better adjust to the external or implanted device. The evolution of augmented intelligence may eventually lead to direct memory sharing between people, chip-directed learning, and even partial immortality (Tipler, 1994; (Koene, 2012) (e.g., by using pre-programmed external devices). Augmented intelligence is a controversial technology because it may easily violate existing ethical norms (Wikipedia contributors, 2012b). In particular, it should not do any harm to people, or violate people's right for privacy and personal identity. Also, developers of this technology have to share partial responsibility for the actions of the customers who use it. Thus, augmented intelligence would require strict regulations from society.

In summary, the functional complexity of human civilization grows exponentially with a doubling time ca. 20 years, but we do not see any signs of an approaching "technological singularity" when humans would be replaced by intelligent machines. Instead, we expect a stronger integration of human mind with technology that would result in augmented intelligence. Creation of new technologies, i.e., emergences, is the norm in the evolution of life (Reid, 2007), and we should not be afraid of it. Our technologies represent the functional envelope for human society, just as intracellular molecular machines make the functional envelope for the DNA molecule. Can we anticipate further reduction of the doubling time in the growth of functional complexity? There is no doubt that exponential "acceleration of returns" will continue because of the positive feedback from increasing functional complexity of human civilization. However, the doubling time, which is inverse of the exponential parameter, is likely to remain stable as it was stable for billions of years in the evolution of the genome. We have already passed through a period of hyperexponential acceleration caused by advances in science, computer technology, Internet, and molecular biology. New technologies will keep emerging but they are unlikely to change the established doubling time for human civilization. There are too many factors that counteract the growth of technology, such as negative population growth in most developed countries, increasing unemployment, unsolved environmental problems, and threats of war and social unrest. Thus, a simple exponential growth seems to be the most realistic forecast.

**10. A Biosemiotic Perspective**

One way to look at the phenomenon of complexity increase is through the eyes of the growing discipline of biosemiotics, in which organisms are considered to be active "agents". Some kind of memory is necessary for self-communication within an agent to preserve its functions. For example, if we learn how to use a hammer and nails, we establish a message within our brain to our future self, which will help us to replicate our actions in the future. In the same way, organisms supply their descendents with genetic "memory" that helps subsequent generations of organisms to replicate their functions. The elements of memory can be called "signs" because they play a role similar to words in a human language. The only difference is that the meaning of words is established via learning and culture, whereas the meaning of genetic signs is heritable.

The natural sciences nowadays tend to deliberately avoid any talk of goals and meanings. Numerous terms have been invented as substitutes, e.g., teleonomy (de Laguna, 1962) or cybernetics (Ashby, 1963). Nevertheless, we deem it necessary to have some straight talk about goals and meanings. Goals in an agent can be self-generated, but they can also be externally programmed by parental agents or higher-level agents. Meanings are stable responses of agents to certain signs. However, the stability of meanings can be supported by various processes from social interactions to assorted heritable functions. Each function supports the meaning of signs that encode other functions. For example, the meaning of coding genome regions is supported by the function of RNA polymerase and DNA polymerase, and the meaning of regulatory genome regions is supported by the function of various transcription factors. Because cellular functions establish meanings for each other, they form a semiotic closure (Pattee, 1995; Joslyn, 2000).

Because of the semiotic nature of life we have to reconsider the notion of biological evolution. The most fundamental process in evolution is accumulation of novel functions which give organisms tools and methods for more successful survival, reproduction, programming of their offspring, capturing resources, and recruiting or reprogramming other agents for their benefit. This process is accompanied by changes of the genome and morphology, but these are observables rather than the core process. Some functions are designed to support morphological plasticity of organisms so that morphology can be easily changed on demand (West-Eberhard, 2002). Thus, dramatic changes in morphology and behavior may occur with very little or even no change in the genome.

The meaning of the DNA depends on the presence of numerous subagents within the cell (e.g., RNA-and DNA-polymerases, ribosomes, and transcription factors). Although manufacturing of these subagents is encoded in some portion of the DNA, the cell requires a minimal number of physically-existing subagents to interpret the DNA. Thus, organisms are not just digital, they require material subagents to process these digital signs. But these additional factors do not challenge the importance of the genome as the major source of heritable information; an assumption that allowed us to trace the rates of progressive evolution and estimate the age of life.